\documentclass[a4paper,UKenglish]{dagrep}

\usepackage{microtype}

\usepackage{url}

\title{
Exploring the Role of Security Experience and ChatGPT Usage Strategies on Secure Software Engineering Education
}
\titlerunning{Role of Security Experience and ChatGPT Usage on Secure Software Engineering}

\author[1]{Alessio Ferrari}
\author[2]{Minh An Nguyen}
\author[2]{Kushal Ramkumar}
\author[2]{Liliana Pasquale}
\affil[1]{Trinity College Dublin (TCD), School of Computer Science and Statistics, Ireland, email: 
  \texttt{aferrari@tcd.ie}}
\affil[2]{University College Dublin, School of Computer Science, Ireland \\ email: \texttt{minh.a.nguyen@ucdconnect.ie,kushal.ramkumar@ucdconnect.ie,\\liliana.pasquale@ucd.ie}}
\authorrunning{A. Ferrari et al.}

\begin{document}

\maketitle

\begin{abstract}
The rapid adoption of Large Language Models (LLMs) is reshaping software engineering education, but their role in secure software engineering education remains underexplored. We report an exploratory empirical study of how 26 graduate students in a part-time MSc Cybersecurity programme used ChatGPT during a vulnerability-fixing assignment. To characterise ChatGPT use, we analysed students' ChatGPT interaction logs using a structured double-coding procedure and examined whether usage patterns and prior cybersecurity expertise were associated with assignment performance.
The results show that students with varying levels of cybersecurity expertise used broadly similar ChatGPT strategies. Individual usage patterns showed descriptive differences by grade, but none remained statistically significant after correcting for multiple comparisons. In contrast, diversity of ChatGPT usage ---the number of distinct usage patterns adopted---was positively associated with performance, even after controlling for cybersecurity expertise. These exploratory findings suggest that the way students engage with ChatGPT may be more informative than whether they use it, and motivate future controlled studies to guide students toward effective LLM use in secure software engineering education.


\end{abstract}

\section{Introduction}
\label{sec:intro}

Generative AI is reshaping programming and software engineering education. LLMs such as ChatGPT can generate code, explain concepts, assist with debugging, and provide immediate feedback, enabling students to seek help while analysing problems, developing solutions, and revising code. At the same time, their use raises concerns about learning quality, assessment, teaching integration, and over-reliance~\cite{agbo2025computing}. These concerns are amplified in secure software engineering modules, where students are expected not only to write functional code, but also to reason about attack vectors and implement mitigations that reduce risk without introducing new weaknesses. A generated solution that passes functional tests may still be insecure, making the use of LLMs in vulnerability fixing potentially risky when students lack the expertise to evaluate generated fixes critically.

Prior work has studied LLM support for programming and software engineering education from several perspectives: code generation and assessment implications~\cite{kiesler2023large}, students' ChatGPT-facilitated programming behaviours and perceptions~\cite{sun2024would}, the effect of ChatGPT access on introductory programming learning~\cite{xue2024does}, and AI tutoring in software engineering education~\cite{frankford2024ai}. Other studies have examined the security of AI-generated code~\cite{pearce2025asleep} and the use of LLMs for vulnerability detection and repair~\cite{fu2023chatgpt,kulsum2024case}. However, prior work has paid less attention to LLM use in secure software engineering education, especially to how different interaction patterns relate to student outcomes.

In this paper, we investigate how students use ChatGPT during cybersecurity vulnerability-fixing activities and how different forms of use relate to task performance. Rather than treating ChatGPT use as a binary variable, we analyse students' interaction logs to identify distinct usage patterns, such as asking for help with implementing specific security controls, refining generated code, debugging errors, exploring alternative solutions, and copying generated code without careful review.

We conducted an exploratory study with 26 participants who used ChatGPT while completing a cybersecurity vulnerability-fixing exercise. Participants had varying levels of cybersecurity expertise, which allowed us to examine whether observed performance differences were associated with prior expertise, specific ChatGPT usage patterns, or the diversity of patterns adopted. We coded students' ChatGPT interactions and analysed both individual usage patterns and the total number of distinct patterns used.

Our results show that students with different cybersecurity expertise levels used broadly similar ChatGPT strategies. Several task-oriented patterns, such as asking for help with a specific security control and refining AI-generated code, were descriptively associated with higher grades. However, no individual pattern remained statistically significant after correcting for multiple comparisons, reflecting the exploratory nature of the study and the limited sample size.

More consistently, usage diversity was positively associated with performance: participants who used a larger number of distinct ChatGPT patterns tended to obtain higher grades. This association remained visible in an exploratory regression model controlling for cybersecurity expertise, while cybersecurity expertise itself was not a statistically significant predictor. These findings suggest that, in this setting, how students engaged with ChatGPT may be more informative than whether they used it, and potentially more informative than prior cybersecurity expertise alone.

We do not interpret these findings as causal evidence. Instead, this study provides an initial empirical characterisation of ChatGPT-supported vulnerability-fixing practices and identifies hypotheses for future controlled studies. In particular, our results motivate further investigation into whether diversified, task-oriented, and reflective use of ChatGPT can help students engage more effectively with secure software engineering tasks.
\section{Related Work}
\label{sec:related}

Prior work has examined the use of LLMs in programming and software engineering education from several perspectives. Kiesler and Schiffner~\cite{kiesler2023large} evaluated ChatGPT-3.5 and GPT-4 on introductory Python tasks, showing that LLMs can solve many standard programming exercises and produce explanations, with implications for assessment design. Sun et al.~\cite{sun2024would} studied ChatGPT-facilitated programming behaviours, performance, and perceptions among college students. In software engineering education, Xue et al.~\cite{xue2024does} found that access to ChatGPT did not necessarily improve learning in an introductory programming course, while Frankford et al.~\cite{frankford2024ai} integrated a GPT-based tutor into an automated programming assessment system and analysed how students interacted with it. These studies show both the potential of LLM-based learning support and the need to understand how students engage with such tools, rather than only whether they used them.

Recent empirical software engineering work has moved toward a more behavioural view of LLM use. Xiao et al.~\cite{xiao2024devgpt} introduced a dataset of developer--ChatGPT conversations linked to software development artefacts, showing that developers use ChatGPT for activities such as code generation, explanation, debugging, documentation, and problem solving. In educational settings, Rahe and Maalej~\cite{rahe2025programming} studied how students use ChatGPT during code-understanding and code-improvement tasks, highlighting risks of reduced programmer agency when students rely on generated solutions without sufficient understanding or revision. Similarly, Hak et al.~\cite{hak2025observing} describe pseudo-apprenticeship patterns in student LLM use, where learners observe or adapt expert-like AI-generated solutions without fully engaging in the reasoning needed to develop independent competence. These studies motivate analysing LLM use as a set of interaction patterns rather than as a single treatment condition.

LLMs have also been studied in security-relevant software engineering tasks. 
Fu et al.~\cite{fu2023chatgpt} evaluated ChatGPT for vulnerability detection, classification, severity estimation, and repair, finding that vulnerability-related tasks remain challenging and often require domain expertise. Kulsum et al.~\cite{kulsum2024case} further showed that reasoning and patch-validation feedback can improve LLM-based vulnerability repair, suggesting that LLMs are most useful when embedded in an iterative validation process. This work shows that LLMs can support security-related programming tasks, but their outputs also require critical evaluation. Our work complements these studies by focusing not on the standalone performance of LLMs on security-relevant software engineering tasks, but on how students use ChatGPT during a vulnerability-fixing exercise, and whether different usage patterns and usage diversity are associated with task performance.
\section{Research Design}
\label{sec:design}

The objective of this study is to characterise students' use of ChatGPT during a vulnerability-fixing assignment and to explore how this use relates to task performance. To this end, we analyse students' ChatGPT conversations to identify usage patterns and measure diversity of use. We also account for prior cybersecurity expertise, as it may influence both students' performance and their ability to evaluate ChatGPT-generated suggestions.

Following the Goal-Question-Metric (GQM) format, our research goal is to 
\textbf{analyse} ChatGPT-supported vulnerability-fixing activities; 
\textbf{for the purpose of} understanding and evaluating students' use of LLM assistance; 
\textbf{with respect to} usage patterns, usage diversity, and vulnerability-fixing performance; 
\textbf{from the viewpoint of} researchers and cybersecurity educators; 
\textbf{in the context of} students with varying cybersecurity expertise levels. We decompose this objective into the following RQs:

\begin{itemize}
    \item \textbf{RQ1:} What ChatGPT usage patterns were most frequently observed, and do these patterns differ descriptively by cybersecurity expertise group?
    
    \item \textbf{RQ2:} Are individual ChatGPT usage patterns associated with differences in vulnerability-fixing performance?

\item \textbf{RQ3:} Is ChatGPT usage diversity, measured as the number of distinct usage patterns, associated with vulnerability-fixing performance?

\item \textbf{RQ4:} Does the relationship between ChatGPT usage diversity and vulnerability-fixing performance remain visible when accounting for cybersecurity expertise?
\end{itemize}

\subsection{Data Collection Procedure}
Participants were opportunistically recruited from a part-time postgraduate Secure Software Engineering module at Anonymous University\footnote{Ethical approval was obtained from the relevant institutional ethics committee prior to data collection. The reference contains the surname of one of the authors and will be disclosed upon paper acceptance to preserve anonymity.}. The analysed subset comprised the $26$ participants who used ChatGPT during the vulnerability-fixing assignment. We collected variables capturing task performance, prior cybersecurity expertise, usage patterns, and usage diversity, as summarised in Table~\ref{tab:variables}.

\begin{table}[htbp]
\centering
\caption{Variables used in the exploratory analysis.}
\label{tab:variables}

\begin{tabular}{|p{0.22\linewidth}|  p{0.7\linewidth}|}

\hline
\textbf{Variable} & \textbf{Description} \\
\hline
\texttt{grade} 
& {Instructor-assigned score for the vulnerability-fixing activity, on a scale $[1-100]$, used as the operational measure of vulnerability-fixing performance.} \\\hline

\texttt{cybersecurity\_exp} 
& Participant's prior cybersecurity expertise, measured on a scale $[0-5]$ through a pre-test and used to derive the cybersecurity expertise group.\\\hline

\texttt{cyber\_exp\_group} 
& Recoded cybersecurity expertise level, distinguishing participants with \emph{medium-low} ($[0-3]$) and \emph{high} ($(3,5]$) cybersecurity expertise. \\\hline

\texttt{P1}--\texttt{P14} 
& ChatGPT usage-pattern indicators, denoting the presence of each ChatGPT usage pattern for a participant. \\\hline

\texttt{n\_patterns\_used} 
& Number of distinct ChatGPT usage patterns observed for a participant, used as a measure of usage diversity. \\\hline
\end{tabular}
\end{table}

Data collection involved a pre-test questionnaire and a vulnerability-fixing assignment.

\textbf{Pre-test questionnaire.} During the first lecture, participants completed a questionnaire collecting demographic information, professional background, and responses to nine cybersecurity questions used to derive prior cybersecurity expertise. Most participants identified as male (88.5\%), held a computer science degree (65.4\%), and worked in the technology sector (80.8\%). Professional experience varied, with 9 participants (34.6\%) reporting more than 10 years of experience. To account for the imbalance in pre-test cybersecurity expertise, participants were classified into two groups: medium-low expertise ($[0, 3]$; $9$ participants, $34.6\%$) and high expertise ($(3, 5]$; $17$ participants, $65.4\%$). 

\textbf{Vulnerability-fixing assignment.} The assignment required students to improve the security of a web application by fixing vulnerabilities identified in a peer-generated report and in a predefined list based on the OWASP Top 10 2021 and associated CWE entries. Students submitted the application together with a report describing each fixed vulnerability, its location, the selected fix, and the implemented security control. Students who used ChatGPT also completed a short questionnaire about their use of the tool and submitted their complete ChatGPT conversation history.


\subsection{Data Analysis Procedure}

The analysis was exploratory and aimed to identify descriptive patterns and generate hypotheses for future confirmatory studies. We therefore emphasise the direction, magnitude, and consistency of effects across analyses rather than treating statistical significance as definitive evidence. A mapping between the research questions and the analysis procedure is provided in Table~\ref{tab:rq_analysis}

To address RQ1, two researchers independently coded participants' ChatGPT conversations to identify usage patterns and assess coding reliability. The coding followed a semi-open procedure informed by Saldaña~\cite{saldana2021coding}. An initial set of codes was defined by the authors based on the study objectives and expected forms of ChatGPT use; additional codes were added when new patterns emerged during coding. Any disagreements were discussed and resolved through consensus. The final codes were then transformed into binary indicators, \texttt{P1}--\texttt{P14}, denoting whether each pattern was observed for each participant. We then computed the frequency of each pattern overall and separately by cybersecurity expertise group.

 To address RQ2, we compared vulnerability-fixing performance between participants who used each individual pattern and those who did not. These comparisons were primarily descriptive. Where appropriate, we also used exploratory Wilcoxon rank-sum tests to compare grade distributions between users and non-users of each pattern. Because multiple pattern-level comparisons were conducted, the resulting p-values were interpreted cautiously and used only as exploratory indicators.

To address RQ3, we computed \texttt{n\_patterns\_used}, defined as the number of distinct ChatGPT usage patterns observed for each participant. This variable was used as a measure of diversity in ChatGPT usage. We examined its association with vulnerability-fixing performance using visual inspection and exploratory regression analysis. 

To address RQ4, we estimated an exploratory linear regression model with \texttt{grade} as the outcome, \texttt{n\_patterns\_used} as the main predictor, and cybersecurity expertise group as a control variable. This model was used to examine whether the association between usage diversity and vulnerability-fixing performance remained visible after accounting for prior cybersecurity expertise. Regression assumptions were inspected using standard diagnostic plots. Given the small sample size and observational design, we do not make causal claims from these analyses.

\begin{table}[t]
\centering
\caption{Mapping between research questions and analysis procedures.}
\label{tab:rq_analysis}
\begin{tabular}{|p{0.06\linewidth} | p{0.85\linewidth}|}
\hline
\textbf{RQ} & \textbf{Analysis procedure} \\
\hline
RQ1 
& Frequency counts of each usage pattern (\texttt{P1}--\texttt{P14}) overall and separately by cybersecurity expertise group (\texttt{cyber\_exp\_group}). \\\hline

RQ2 
& Descriptive comparison of grades between participants who used each pattern and those who did not; exploratory Wilcoxon rank-sum comparisons where appropriate. \\\hline

RQ3 
& Scatterplot and exploratory linear regression examining the association between pattern usage diversity (\texttt{n\_patterns\_used}) and grade. \\\hline

RQ4 

& Exploratory linear regression predicting grade from usage diversity while controlling for the cybersecurity expertise group. \\
\hline
\end{tabular}
\end{table}
\section{Results}
\label{sec:results}

\subsection{RQ1: Distribution of ChatGPT Usage Patterns}

\begin{figure}
    \centering
    \includegraphics[width=1\linewidth]{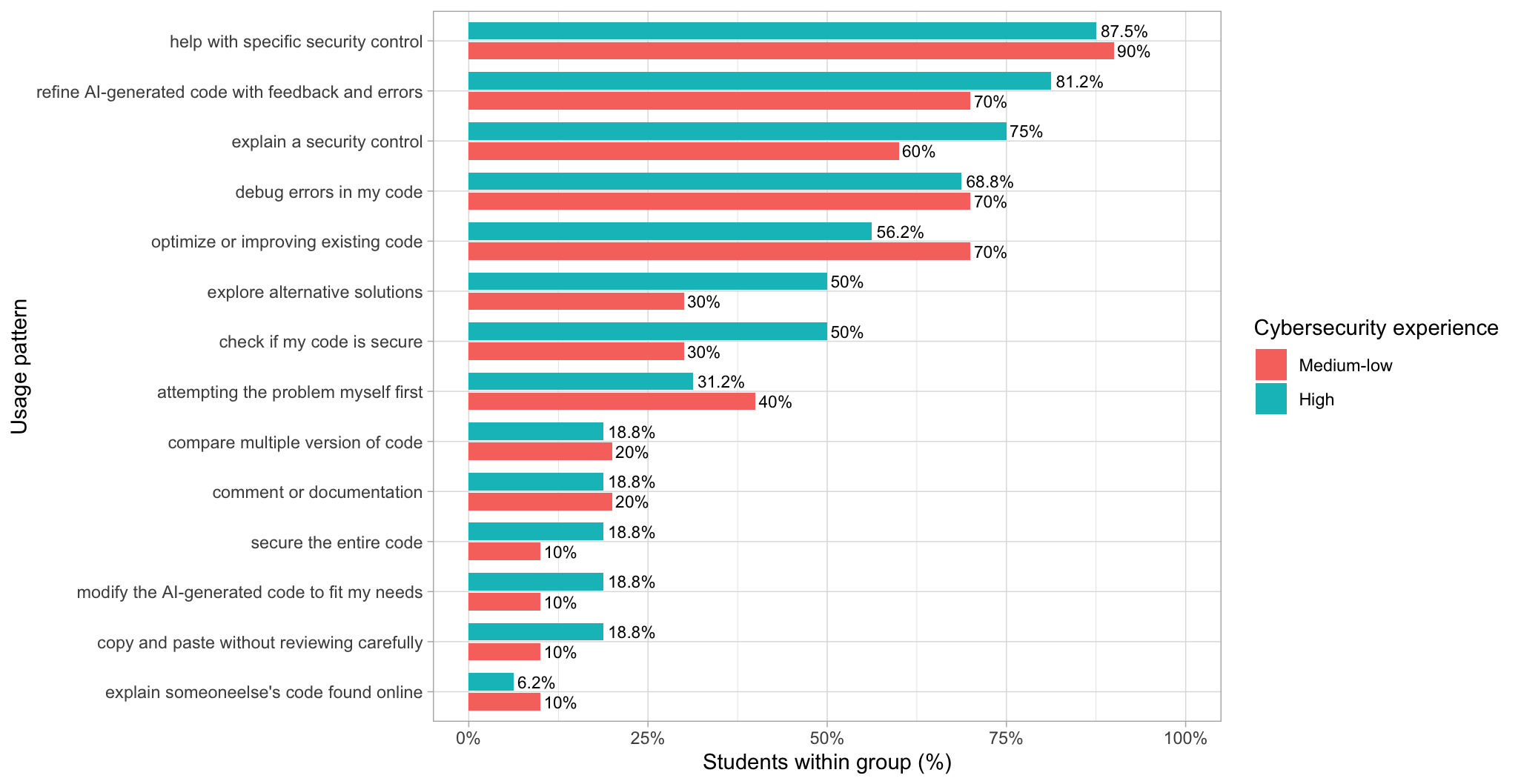}
    \caption{Frequency of pattern usage by cybersecurity expertise group.}
    \label{fig:patterns_by_expertise}
\end{figure}

Figure~\ref{fig:patterns_by_expertise} shows the distribution of ChatGPT usage patterns among participants grouped by cybersecurity expertise. Percentages are computed within each expertise group.

The most frequent patterns were broadly similar across groups. In both groups, the dominant pattern was asking ChatGPT to generate an implementation for a specific mitigation, observed for 90\% of participants in the medium-low expertise group and 87.5\% in the high expertise group. Other common patterns included refining AI-generated code using feedback or errors, asking for explanations of security controls, debugging code, and optimising or improving existing code.

Overall, the distribution does not suggest a clear separation between medium-low and high cybersecurity expertise participants. Some descriptive differences can nevertheless be observed. Participants in the high expertise group more often used ChatGPT to refine generated code, explain security controls, explore alternative solutions, and check whether their code was secure. Participants in the medium-low group more often used ChatGPT to optimise existing code and to attempt the problem themselves before relying on ChatGPT. Less frequent behaviours, such as comparing multiple versions of code, adding comments or documentation, securing the entire application, adapting generated code to specific needs, and copying code without careful review, appeared only for a minority of participants.

These differences should be interpreted cautiously. The expertise groups were small and unbalanced, with 10 participants in the medium-low group and 16 in the high group. As a result, small changes in participant counts correspond to relatively large percentage changes. We therefore treat the observed differences as descriptive patterns rather than evidence of systematic differences between expertise groups.

In summary, RQ1 suggests that participants with different levels of cybersecurity expertise used broadly similar ChatGPT strategies. Both groups primarily used ChatGPT for direct task support, including generating and understanding specific security controls, refining code, and debugging errors. We next examine whether these usage patterns are associated with vulnerability-fixing performance.

\subsection{RQ2: Usage Patterns and Observed Grade Differences}

\begin{figure}
\centering
\includegraphics[width=0.65\linewidth]{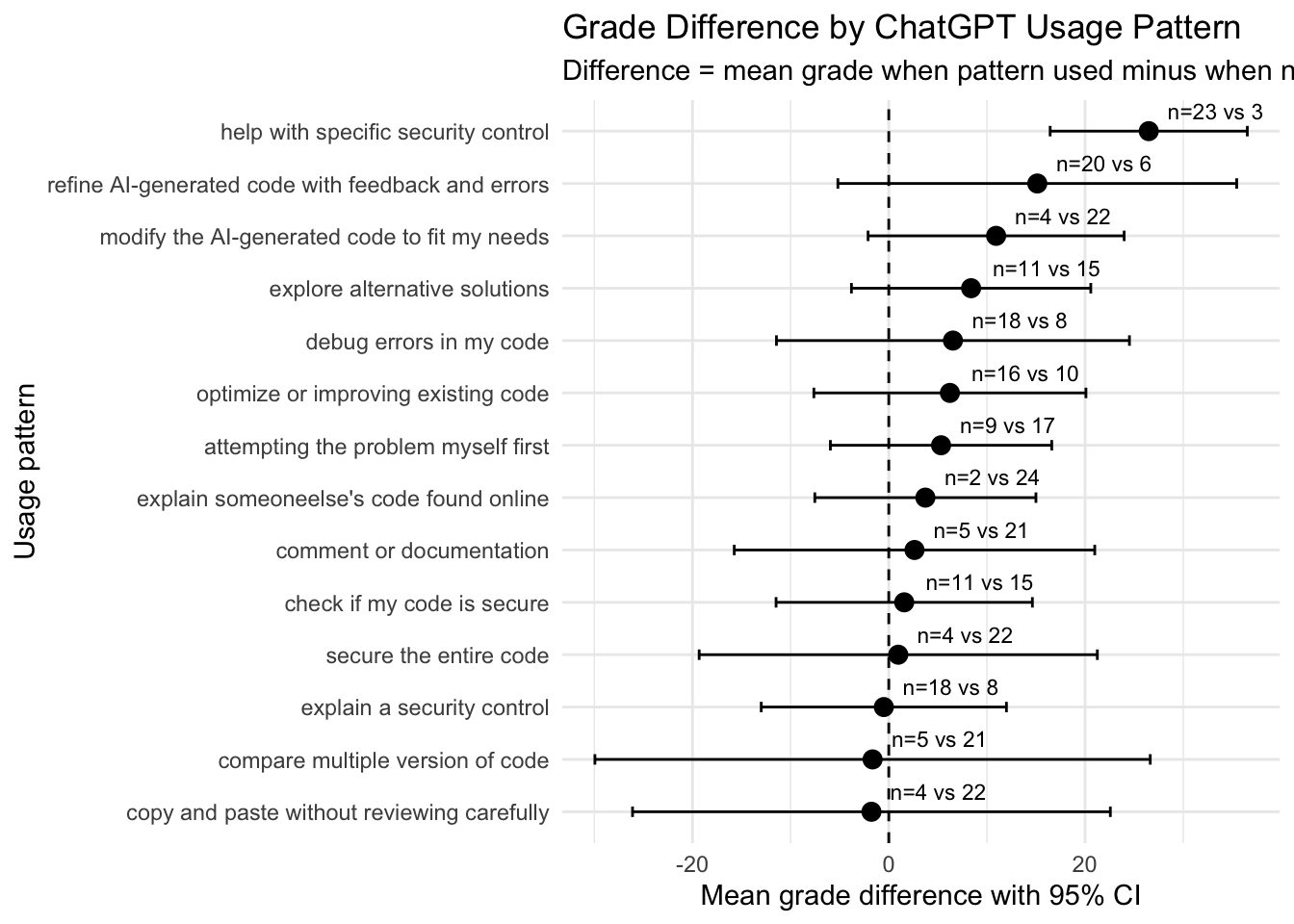}
\caption{Grade difference by ChatGPT usage pattern.}
\label{fig:pattern_grade_differences}
\end{figure}

RQ2 examined whether individual ChatGPT usage patterns were associated with differences in vulnerability-fixing performance. Figure~\ref{fig:pattern_grade_differences} reports, for each pattern, the difference between the mean grade of participants who used the pattern and those who did not. Positive values indicate higher average grades among participants who used the pattern.

Most usage patterns were associated with positive mean grade differences. The largest difference was observed for \emph{generate code based on mitigation name} ($+26.5$ points), although this pattern was highly imbalanced, with 23 participants using it and only 3 not using it. Other positive differences were observed for \emph{refine AI-generated code with feedback and errors} ($+15.1$), \emph{modify the AI-generated code to fit my needs} ($+10.9$), and \emph{explore alternative solutions} ($+8.39$). Smaller positive differences were observed for patterns such as debugging errors, optimising existing code, and attempting the problem before relying on ChatGPT.

Only a few patterns showed near-zero or negative differences. These included \emph{copy and paste without reviewing carefully} ($-1.77$), \emph{compare multiple versions of code} ($-1.65$), and \emph{explain a security control} ($-0.50$). The negative value for copying without review is consistent with the interpretation that less reflective ChatGPT use was not associated with better performance, but this pattern was observed for only four participants and should be interpreted cautiously.

We complemented these descriptive comparisons with exploratory Wilcoxon rank-sum tests. Before correction, only \emph{generate code based on mitigation name} showed a statistically significant difference ($p=.016$). After applying the Benjamini--Hochberg correction for multiple comparisons, no individual pattern remained statistically significant. Thus, while several patterns were descriptively associated with higher grades, the evidence for any single pattern was not statistically robust. This motivates the next analysis, which examines whether overall usage diversity is associated with vulnerability-fixing performance.

\subsection{RQ3: Usage Diversity and Grade}

\begin{figure}
    \centering
    \includegraphics[width=0.7\linewidth]{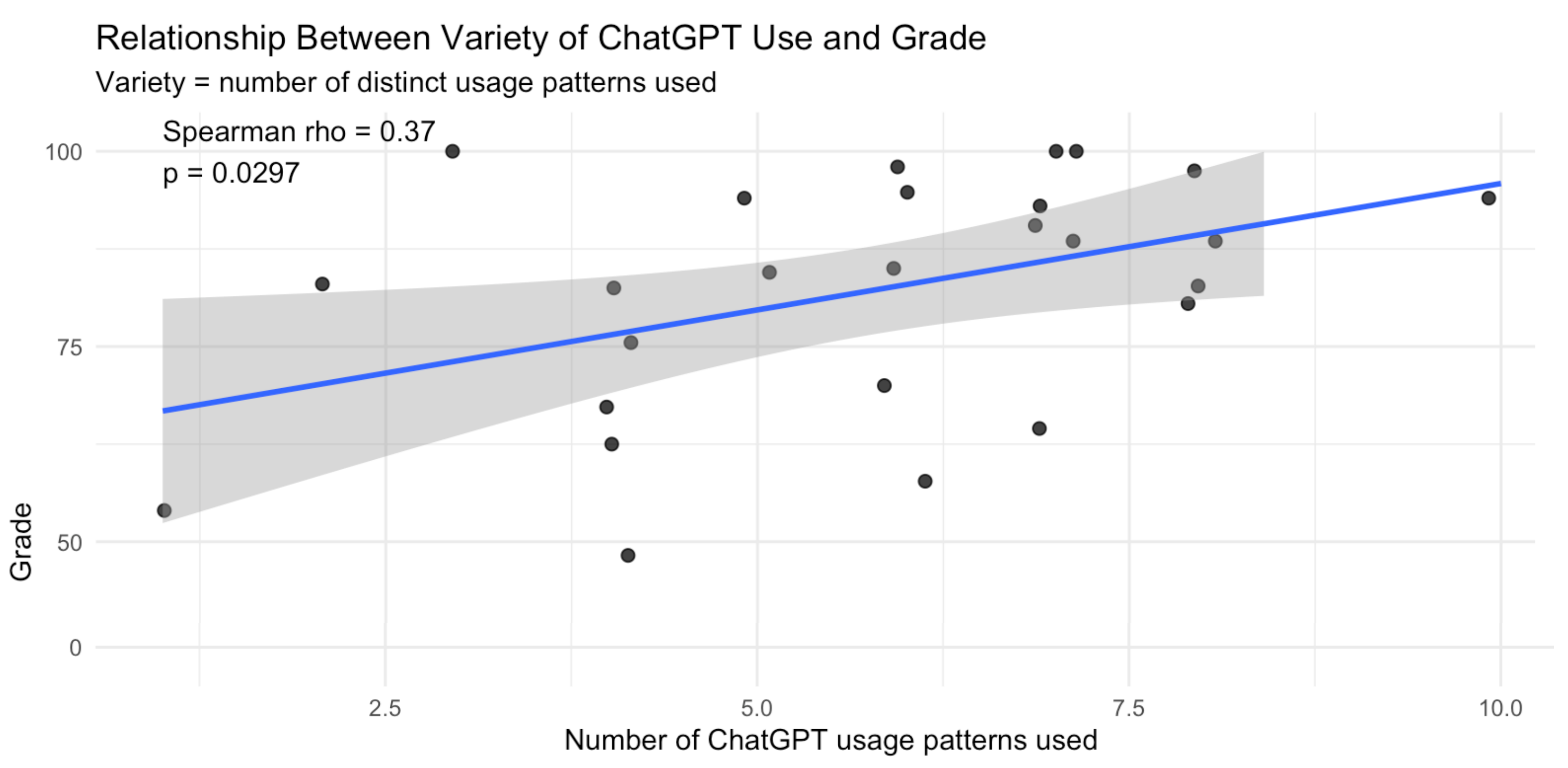}
    \caption{Linear model showing the relationship between number of patterns used and grade.}
    \label{fig:variety_grade}
\end{figure}

RQ3 examined whether ChatGPT usage diversity was associated with vulnerability-fixing performance,  operationalised as \texttt{grade}. Usage diversity was measured as the number of distinct ChatGPT usage patterns observed for each participant. As shown in Figure~\ref{fig:variety_grade}, participants who used a wider range of patterns tended to obtain higher grades.

A one-sided Spearman rank correlation showed a moderate positive association between usage diversity and grade ($\rho = .375$, $p = .030$). This suggests that broader engagement with ChatGPT was associated with better task performance. However, the scatterplot also shows substantial variability, indicating that usage diversity explains only part of the observed differences in grades.

Overall, RQ3 suggests that students who engaged with ChatGPT through a wider range of usage patterns tended to perform better. RQ4 therefore examines whether this association remains visible after accounting for participants' cybersecurity expertise.

\subsection{RQ4: Usage Diversity, Grade, and Cybersecurity Expertise}

RQ4 examined whether the association between ChatGPT usage diversity and vulnerability-fixing performance remained visible after accounting for cybersecurity expertise. To test this, we fitted an exploratory linear regression model with \texttt{grade} as the outcome, \texttt{n\_patterns\_used} as the main predictor, and \texttt{cyber\_exp\_group} as a control variable, coded as $0$ for medium-low expertise and $1$ for high expertise:

{\footnotesize
\[
\widehat{grade} = 61.95 + 3.11 \times \texttt{n\_patterns\_used} + 3.71 \times \texttt{cyber\_exp\_group}
\]
}

The coefficient for usage diversity was positive, $b = 3.11$, $SE = 1.37$, $p = .033$, 95\% CI $[0.28, 5.95]$, indicating that each additional ChatGPT usage pattern was associated with an estimated increase of 3.11 grade points, holding cybersecurity expertise constant. In contrast, cybersecurity expertise group was not clearly associated with grade, $b = 3.71$, $SE = 5.76$, $p = .526$, 95\% CI $[-8.20, 15.62]$. The model explained approximately 21\% of the variance in grades ($R^2 = .21$), although the overall model test did not reach the conventional .05 threshold, $F(2, 23) = 3.06$, $p = .067$. Standard diagnostic plots did not suggest major violations of linear regression assumptions, and the Shapiro--Wilk test did not indicate a strong departure from normality ($p = .785$). Overall, the results suggest that usage diversity remained positively associated with grade after accounting for cybersecurity expertise, but the small sample size and exploratory design warrant cautious interpretation.

\section{Discussion}
\label{sec:discussion}


Our exploratory results suggest that students' vulnerability-fixing performance was associated less with any single ChatGPT usage pattern than with the breadth of their engagement with the tool. Although several individual patterns showed descriptive differences in grade, none remained statistically significant after correcting for multiple comparisons. In contrast, usage diversity showed a more consistent positive association with performance, including when controlling for prior cybersecurity expertise. This suggests that students who used ChatGPT in multiple complementary ways---for example, to generate candidate fixes, debug errors, refine code, and explore alternatives---may have been better able to incorporate the tool into an active problem-solving process. These findings align with prior work showing that access to ChatGPT alone does not determine learning outcomes~\cite{xue2024does}.

For educators and practitioners involved in secure software engineering training, the findings suggest that simply allowing or prohibiting the use of ChatGPT may be less productive than teaching learners to use it critically. In secure software engineering modules, students could be guided to use ChatGPT as a support tool for understanding, debugging, refining, and validating fixes, rather than as a source of code to copy. In practice, this could involve modelling interaction patterns such as asking targeted questions about security controls, requesting explanations of proposed fixes, checking whether generated code addresses the vulnerability, and comparing alternative mitigations. Assessment criteria could also reward evidence of critical engagement with LLM outputs, for example, by asking students to document how generated suggestions were evaluated, adapted, and tested. This is particularly important in security contexts, where a fix that appears functionally correct may still fail to remove the vulnerability or may introduce new weaknesses. Such guidance may be especially valuable for students or early-career practitioners with limited prior cybersecurity expertise.

Because our study is exploratory, these implications should be treated as hypotheses rather than conclusions. For empirical software engineering research, the results motivate moving beyond binary measures of LLM use: treating ChatGPT use as simply present or absent may hide important differences in students' interaction strategies. Future studies should therefore capture interaction-level data, such as chat logs, prompts, revisions, and verification behaviours, and test whether students can be guided toward more effective forms of interaction. Table~\ref{tab:hypothesis} summarises the main hypotheses that follow from our findings. H1 and H2 directly reflect the observed association between usage diversity, vulnerability-fixing performance, and cybersecurity expertise. H3 translates this association into a possible instructional intervention. H4 captures the contrast between reflective, task-oriented use and passive copying, while H5 concerns the possibility that students with different levels of cybersecurity expertise benefit from different forms of ChatGPT support.

\begin{table}[t]
\centering
\caption{Hypotheses for future controlled studies.}
\label{tab:hypothesis}
\begin{tabular}{p{0.03\linewidth} p{0.54\linewidth} p{0.35\linewidth}}
\hline
\textbf{ID} & \textbf{Hypothesis} & \textbf{Rationale} \\
\hline

H1
& Students who use a more diverse set of patterns will achieve higher vulnerability-fixing performance.
& Based on the observed positive association between usage diversity and grade. \\\hline

H2
& The association between ChatGPT usage diversity and vulnerability-fixing performance will remain positive after accounting for prior cybersecurity expertise.
& Based on the exploratory regression model controlling for cybersecurity expertise. \\\hline

H3
& Students guided to use multiple complementary ChatGPT usage patterns will outperform students who use ChatGPT without such guidance.
& Motivated by the need to translate the observed association into a controlled intervention. \\\hline

H4
& Reflective and task-oriented ChatGPT use will lead to better vulnerability-fixing performance than passive use.
& Motivated by the descriptive contrast between task-oriented patterns and copy-paste behaviour. \\\hline

H5
& The benefits of ChatGPT usage patterns will vary depending on students' prior cybersecurity expertise.
& Motivated by the possibility that expertise shapes how students evaluate and apply ChatGPT output. \\

\hline
\end{tabular}
\end{table}

\section{Threats to Validity}
\label{sec:threats}

\textbf{Construct validity.} ChatGPT use was operationalised through manually coded usage patterns. While these patterns provide a structured way to describe participants' interactions, they may not capture differences in depth, quality, or critical reflection within the same pattern. Similarly, \texttt{n\_patterns\_used} captures usage diversity but not the effectiveness of each interaction. To reduce subjectivity, the coding was checked by a second researcher. Task performance was measured using the instructor-assigned grade for the vulnerability-fixing activity, which may not capture all aspects of successful fixing, such as code quality, reasoning process, or security robustness.

\noindent
\textbf{Internal validity.} The study is observational, and participants were not randomly assigned to use specific ChatGPT patterns. Students who used more patterns may also differ in motivation, programming ability, persistence, or general problem-solving skill. We controlled for prior cybersecurity expertise, but other unmeasured confounders may remain. In addition, participants often used multiple patterns together, making it difficult to isolate the effect of any single pattern. For this reason, pattern-level comparisons are interpreted descriptively, and no causal claims are made.

\noindent
\textbf{Conclusion validity.} The small sample size limits statistical power, especially because some patterns were used by only a few participants, creating imbalanced comparisons between users and non-users. We therefore emphasise effect direction, confidence intervals, and visual trends rather than relying only on statistical significance.

\noindent
\textbf{External validity.} Participants were students completing a specific vulnerability-fixing activity in a single educational context. The findings may not generalise to professional developers, other courses, or other security tasks. We therefore present the results as exploratory evidence and describe the coding scheme so that future work can compare, refine, or extend it in larger and more diverse settings.
\section{Conclusion}
\label{sec:conclusion}

This study explored how students used ChatGPT during a cybersecurity vulnerability-fixing exercise and how usage patterns related to task performance. Our results suggest that performance was more consistently associated with diversity in ChatGPT usage than with any individual pattern or with prior cybersecurity expertise alone. Students who used ChatGPT in a wider range of ways tended to achieve higher grades, even after accounting for cybersecurity expertise. Given the exploratory design and limited sample size, these findings should be interpreted as hypothesis-generating rather than causal evidence. Future controlled studies should test whether guiding students toward diversified, task-oriented, and reflective ChatGPT use improves vulnerability-fixing performance.

\section*{Data Availability}
The anonymised replication package, including the study data and analysis scripts, is available on Zenodo at DOI: \texttt{10.5281/zenodo.20453130}. The public record will be updated with author-identifying information after the review process.
Note that the study data include the pre-test questionnaire, the assignment description, the rubric used by the instructor to evaluate performance, and an explanation of the codes used with examples.

\bibliography{ref}

\end{document}